# The Architectural Dynamics of Encapsulated Botnet Detection (EDM)

**Maxwell Scale Uwadia Osagie[1*] and Amenze Joy Osagie[1]**

[1]*Department of Physical Sciences, Faculty of Science, Benson Idahosa University, P.M.B. 1100, GRA, Benin City, Edo State, Nigeria.*

## ABSTRACT

Botnet is one of the numerous attacks ravaging the networking environment. Its approach is said to be brutal and dangerous to network infrastructures as well as client systems. Since the introduction of botnet, different design methods have been employed to solve the divergent approach but the method of taking over servers and client systems is unabated. To solve this, we first identify Mpack, ICEpack and Fiesta as enhanced IRC tool. The analysis of its role in data exchange using OSI model was carried out. This further gave the needed proposal to the development of a High level architecture representing the structural mechanism and the defensive mechanism within network server so as to control the botnet trend. Finally, the architecture was designed to respond in a proactive state when scanning and synergizing the double data verification modules in an encapsulation manner within server system.

## 1. INTRODUCTION

The commercial viability of internet all these years has made people embrace its sustainability and it has become commercial hub, creating thousands of jobs and numerous opportunities to its end users. This hitherto, has increased the scope of the intended idea of the computer and internet landscape and further offer opportunities to ever growing entrepreneurs who uses the platform in creating jobs. The advancement in technology has increased the rate of crime in the system infrastructure network.

_____
*\*Corresponding author: E-mail: mosagie@biu.edu.ng;*

Crime can be said to be the actual act of a criminality. That is, the end product of what has be committed against the law. Crime is in variance with the order agreed upon by the people living in a segmented area that such law protects. The said intended criminal act or the act itself has become nightmare to internet users. A cybercrime, [1] "is a crime committed using the Internet". This involves stealing of personal details or bank details or facilitating connected computers with malwares such as virus and other useful tools that pose serious threat to the functionality of client or server system. A critical examination to what cybercrime represent, means that the crime could be from anywhere. This is the fundamental problem with crime committed online and this is due to the borderless platform restricting individuals who have agreed on some set of rules to guide them in achieving their set goals.

In recent times, some literatures have explained the dichotomy in the literal meaning of the dual hacker and cracker. "Quantity, Supply, Mean" are best captured in economics and there is no other definition outside economics that can best explain these terminologies. Hackers and crackers are computer science terminologies and should be seen from the right perspective when captured by computer science literatures. There is a huge difference to the functionality and personality of hackers and crackers [2].

Hackers are legitimate internet users who have the understanding of the back end workings of a computer as a machine and its functionality in networking environment. Though, crackers are considered to be the same but the method of operation differs. Hackers are the background decoders of computers while crackers are said not to be legitimate users, they can gain access to network without authorization and steal valuable credentials. They use different software tools like botnet in navigating networks so as to exploits vulnerable systems connected. They are known to be harmful and dangerous. In addition, crackers uses vulnerable network from server end to turning and controlling connected client systems [3].

Researchers have showed in recent times how Nigerians have embraced electronic innovations [4]. The use of hand-held devices such as smart phones amongst developing countries populace and the money being made from sales of data bundles by service providers like (MTN, GLO, ETISALAT etc) is a clear indication of how well developing countries like Nigerians have become more responsive to computer literacy [5,6]. There is virtually no citizens of the developing countries without computing gadget and many of them have their gadgets or personal computers (PCs) invaded without authorization. The ideological foundation of smart phones users in this part of the world seems to be quite different from those in the developed countries. While attention is being pay to the front end of such hand-held devices (i.e making of calls and receiving of text messages) little attention is being given to the security mechanism of such hand-held devices. The resultant effect seems to be more devastating than the none attention and time put into the use of the hand-held devices [7,8,9].

This work is structured as follows: Section 2 illustrates botnet operations, section 3 show a high level architecture of EDM, Section 4 gave detail explanation to the components embedded in the EDM, section 4.1 shows section of the navigating approach of botnet propagation and the operational procedure of the architecture.

## 2. RELATED LITERATURES

Several reviews have been carried on botnet tools and techniques. However, before adequate attention can be given to these literatures, detail explanation is given to the movement of data within network space. Fig. 1 represents communication within network. Information is exchange when there is adherence to protocol standard and guidelines rules. Each layer in the Fig. 1 explained the movement of data from computer T to computer P. The movement of data starts by the activation of signal between connection oriented systems because their movement is on the understanding of the workability of the protocols.

A botmaster is an intelligent cracker or invader who specializes in reading every bit of what the protocol movement represents and thereafter uses it against the said goal. Learning how to hack is an interesting thing but required time and skill which could be acquired by constant reading of hacking materials. Hacking has nothing to do with spirituality and if you have no idea of what it entails, there is little or no contribution you can make to secure a data. As it stands, everyone using network is at the mercy of a botmaster.

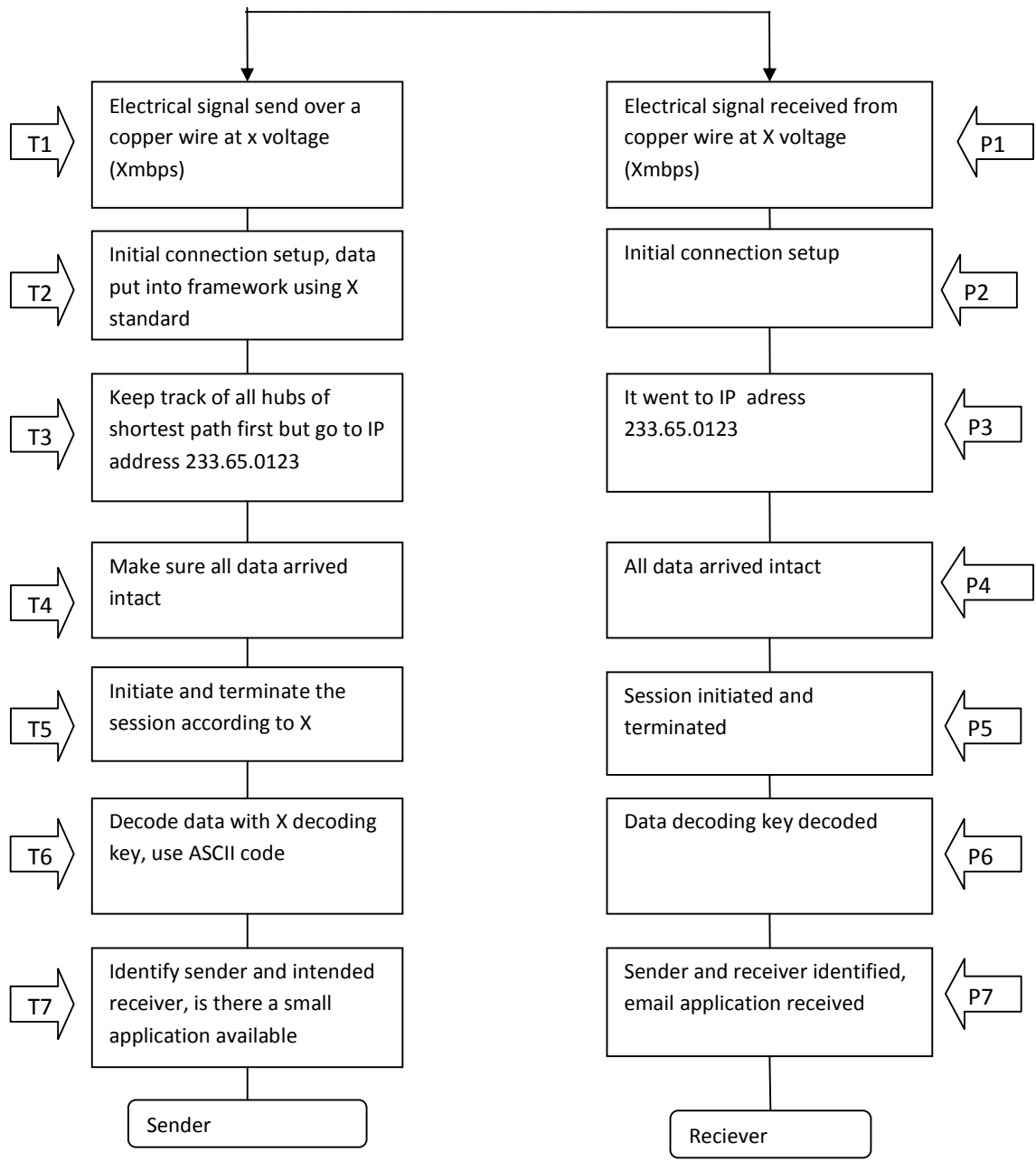

**Fig. 1. Computer communication with adherence to OSI/ISO model standard [10]**

A botmaster is good at studying traffic within network. He does this by observing, scanning and controlling [11,12,13,6]. However, it is difficult to classify them as none legitimate users because they pimp tent with the network link considered to be legitimate [14,6]. There is little or no work to be done on data sharing without the linking of segmented networks. If there is anything that has brought data synergy amongst networking platform, it is the ability for a client system to connect to a server network. Furthermore, a network is an agreed platform where two or more people share resources such as idea, experience, plans etc. but in the case of computer network or networking, it is a connection of two or more systems for resource sharing.

Botmasters resilience to use computing tools in fighting back the state of the system has remained worrisome to server operators. The HTTP has become the new destination where all kinds of messages are sent to intended victim with an awaiting click for command and control (C&C). This actually saw a turnaround from the traditional method to a well robust website platform where HTTP is the protocol for navigating data between website. This is a well matured system for exploitation, in which case, exploiting kits such as Mpack, ICEpack and Fiesta are used. The method uses these M pack, ICE pack and Fiesta in sending messages that could compromise the integrity of connected system. The messages can move the site destination to a more vulnerable environment [15,16,6].

What makes botnet one of the most fearful and dangerous threat in the internet world is the invisible manners at which computers connected to the affected server are invaded without prior knowledge of the owner. Bot-master uses command and control (C&C) channel in activating botnet on server end. The botnet can remain on the server end awaiting a command from the client computer so as to authenticate its operations, with a click on a button on the client-end a signal is sent to the botnet which will eventually make all connected clients bot. Thus, enable the bot-master to have free access in stealing valuables or resource [17,6].

Research has unveiled details about the operational modules of botnet activities. In recent times, mobile botnet has become the destination of bot-master and this is due to the proliferation of internet oriented mobile devices. In a work done by Farina et al. [18] it was revealed that " the development and diffusion of mobile devices such as smart phones with internet accessibility through Wi-Fi, 3G, LTE, etc " has given the needed technique for mobile devices attack. They also proposed in the same article that adequate attention be given to the development of mobile botnet protection. In a related work by Papaleo et al. [19], the rational behind the development of Trojan horse through critical examination of android platform was discussed. The work unveiled the development of a control mechanism called android total control that could mitigate the threat.

## 3. HIGH LEVEL ARCHITECTURE OF ENCAPSULATED DETECTION MECHANISM (EDM)

The operation of botnet tools within network is on the understanding of the principle of client-server connected systems synergy. The client system in Fig. 2 represents any possible users who depend on the services of the server system. The server as represent in the architecture above has double verification mechanism and the database validation. The server is protected by the botnet guide mechanism which does all pre-activities of all entry connection that may be P2P, IRC and HTTP. The botmaster is the unauthorized persons that specialises in the use of botnet tools. Fig. 3 gave details explanation to each segments of the architecture and the possible response to the botmaster's threat (botnet tool). The dynamics surrounding the propagation of the operation of the botnet have made different architecture failed in the quest to protecting server systems from being attack by a botmaster. Nevertheless, the high level architecture is a combination of the idea from other architectures thus far theorised and practiced in securing the integrity of data within server scope.

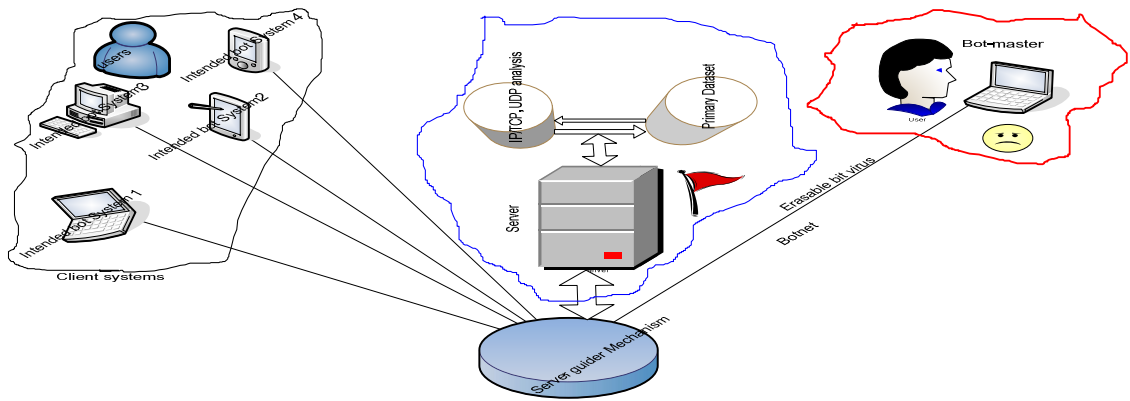

**Fig. 2. Encapsulated detection mechanism (EDM) for Botnet on Server end [6]**

## 4. DETAIL ARCHITECTURE OF AN ENCAPSULATED DETECTION MECHANISM (EDM)

The architectural design as seen in Fig. 3 is of three segments handling (IRC, HTTP and P2P topologies). The segments work as an entity. To solve the botmaster trend, proactive detection and fight back mechanism was developed on server end through exploitation and deployment of several techniques. Fig. 4 below represents design structure in a conceptual frame with an embedded mathematics model (Outlier Analysis) for adequate validation of data within the scope of server. The server guider mechanism is the proposed encapsulated module that serves as gate way to both the client and server systems. The encapsulated mechanism has some dynamic features that help keep the integrity of the server and these are:

i. Fight Back Mechanism Modules (FBMM),
ii. Double Dataset Verification Factor (DDVF),
iii. Captcha Module (CM) and
iv. Three navigation approaches of a Botmaster Propagation (TNABP).

### 4.1 Segment of an Encapsulated Detection Mechanism

According [6] the architecture as shown in Fig. 4 is justified by the embedded encapsulated model divided into three segments, it creates authentication on the entry layer, detection and fight back mechanism on the functional layer of the DM server. The system is designed in three modules.

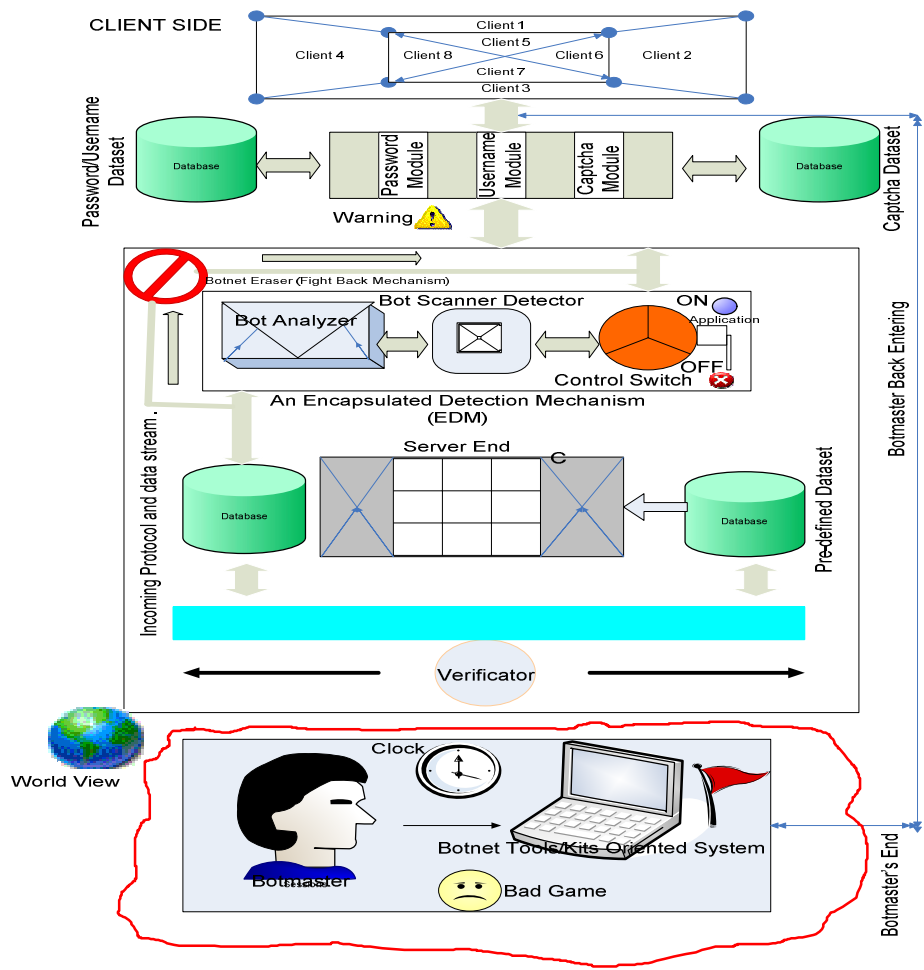

**Fig. 3. Architecture of an encapsulated detection mechanism (EDM) for Botnet on server end [6]**

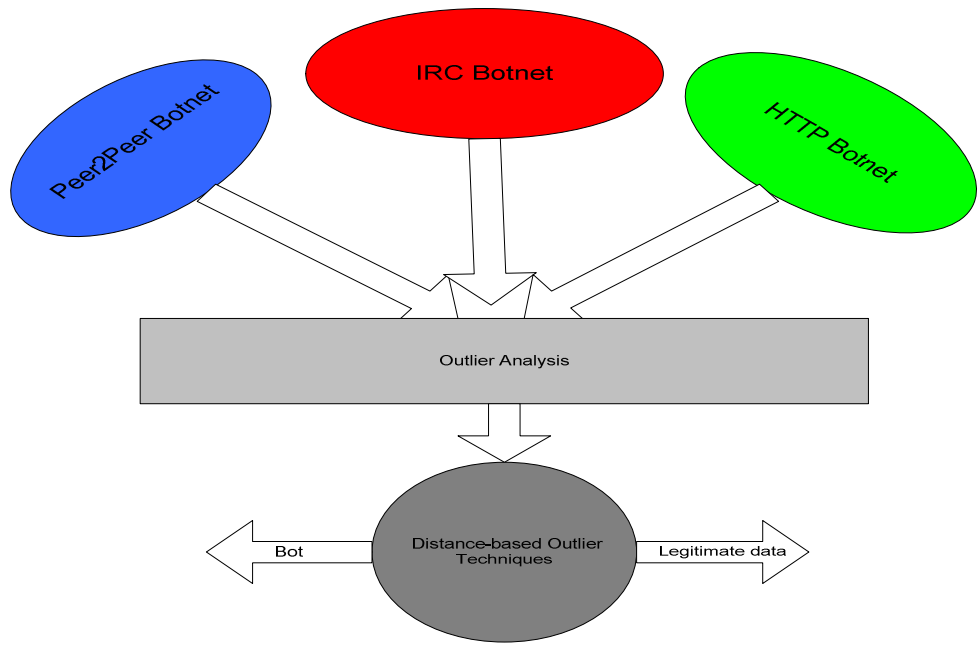

**Fig. 4. Session of the detection mechanism for Botnet on server systems architecture**

**I. User Layer**

This layer has the enduring process of all legitimate users who at a point made and synchronized confidentiality with the system design for onward recognition (handshake /signalling) which is a standard practice for exchange of data [6]

**II. Authentication Layers**

a. Captcha: This eradicates suspicious entry on the system and then creates integrity and assigned privileges to legitimate users on the network server. In recent times, this method has become increasingly appreciated by programmers because it eradicate none human from human
b. Username and Password: the system grants access to predefined registered users via this segment. The use of username and password is a standard practice that cut across computing platforms [6]

**III. Functional Analyzers**

a. Encapsulated Detection Mechanism: it does the analysis of window movement in one dimensional array against predefined dataset within the server domain. It has an embedded bot scanner, analyzer and verification agent that work as an entity to stimulate fight back against data movement short of the normal pattern. The Object Oriented Programming approach gave the idea to the formulation of the entity.
b. Bot Scanner: the scanner is logical in its operation and it goes through input data window, streams as well as check against unwanted data that may have found itself into the system domain by way of futuristic propagation approach. The server has ability to scan double verification of data integrity before granting access
c. Bot Analyzer and Verification Agent: the duo initializes a process for termination and further protocol action on data found to be outside the scope of the predefined dataset [6].

**5. CONCLUSION**

The conceptualization, theoretical and practical perspective of any design is on the fundamental principle of the techniques employed. What makes a solid building is on the design approach. The reinforcement of a structural architecture shows the strength and the quality of the end product. In this work, more details were given to the Encapsulated Detection Mechanism (EDM) architecture because the quality of the proactive defensive mechanism is what gave room to the practicability of taking care of the botnet within server space. This action has helped rekindle the eroded trust in the networking environment and thereby giving income on any single investment done in the form of software and infrastructure as a service. Knowing that client computers are the main targeted system of the botnet tools, the architecture employed combinational defensive mechanisms as demonstrated in Figure 3 and 4 to help resolve all front end entry protocols with coherent understanding of rendering bot oriented system out of working state.

## COMPETING INTERESTS

Authors have declared that no competing interests exist. The products used for this research are commonly and predominantly use products in our area of research and country. There is absolutely no conflict of interest between the authors and producers of the products because we do not intend to use these products as an avenue for any litigation but for the advancement of knowledge. Also, the research was not funded by the producing company rather it was funded by personal efforts of the authors.

*Peer-review history:*
*The peer review history for this paper can be accessed here:*
*http://www.sdiarticle3.com/review-history/46652*